\documentclass[%
 reprint,
superscriptaddress,
 amsmath,amssymb,
pra,
floatfix,
 array,
]{revtex4-1}

\usepackage{graphicx}
\usepackage{dcolumn}
\usepackage{bm}
\usepackage{amsthm}
\usepackage{subfigure}
\usepackage[usenames]{color}

\begin{document}


\title{Ising interaction between capacitively-coupled superconducting flux qubits}
\author{Takahiko Satoh}
 \email{satoh@is.s.u-tokyo.ac.jp}
\affiliation{
 NTT Basic Research Laboratories, 3-1, Morinosato Wakamiya Atsugi-city,
 Kanagawa 243-0198 Japan}
\affiliation{
 Department of Computer Science, Graduate School of Information
 Science and Technology, The University of Tokyo, 7-3-1, Hongo,
 Bunkyo-ku, Tokyo, Japan}
\author{Yuichiro Matsuzaki}
\affiliation{
 NTT Basic Research Laboratories, 3-1, Morinosato Wakamiya Atsugi-city,
 Kanagawa 243-0198 Japan}
\author{Kosuke Kakuyanagi}
\affiliation{
 NTT Basic Research Laboratories, 3-1, Morinosato Wakamiya Atsugi-city,
 Kanagawa 243-0198 Japan}
\author{Koichi Semba}
\affiliation{
 Advanced ICT Research Institute, National Institute of Information and
 Communications Technology, 4-2-1, Nukuikitamachi, Koganei-city, Tokyo
 184-8795 Japan}
\author{Hiroshi Yamaguchi}
\affiliation{
 NTT Basic Research Laboratories, 3-1, Morinosato Wakamiya Atsugi-city,
 Kanagawa 243-0198 Japan}
\author{Shiro Saito}
\affiliation{
 NTT Basic Research Laboratories, 3-1, Morinosato Wakamiya Atsugi-city,
 Kanagawa 243-0198 Japan}

\date{\today}

\begin{abstract}
Here, we propose a scheme to generate a controllable Ising interaction
between superconducting flux qubits.
Existing schemes rely on inducting couplings to realize Ising
interactions between flux qubits, and the interaction strength is
controlled by an applied magnetic field
On the other hand, we have found a way to generate an
interaction between the flux qubits via capacitive couplings. This has
an advantage in individual addressability, because we can control the
interaction strength by changing an applied voltage that can be easily
localized. This is a crucial step toward the realizing superconducting
flux qubit quantum computation. 
\end{abstract}

\maketitle

\section{Introduction}
To realize fault-tolerant quantum computation, it is crucial to investigate a scheme
to generate a cluster state in a scalable way. The cluster state is a
universal resource for quantum computation, and this state can be used
for a fault-tolerant scheme such as a surface code and topological
code. One can generate a cluster state if we can turn on/off an Ising
type interaction between qubits.

Superconducting circuit is one of the promising systems to realize
such a cluster-state quantum computation. Josephson junctions in the
superconducting circuit can induce a non-linearity, and so one can
construct a two-level system.
There are several types of Josephson junction qubit: charge 
qubit~\cite{SQ_charge}, superconducting spin qubit~\cite{SQ_spin},
superconducting flux
qubit~\cite{SQ_flux,SQ_flux_2,SQ_flux_3,SQ_flux_4,SQ_flux_5}, superconducting phase
qubit~\cite{SQ_phase,SQ_phase_2,SQ_phase_3}, superconducting transmon
qubit~\cite{SQ_transmon,SQ_transmon_2}, fluxonium
qubit~\cite{SQ_fluxonium,SQ_fluxonium_2},  and several hybrid
systems~\cite{SQ_hybrid,SQ_hybrid_2}.

The transmon qubit~\cite{SQ_transmon,SQ_transmon_2,SQ_transmon_3}, which is a cooper-pair box and relatively
insensitive to low-frequency charge noise, is considered one of the
powerful method of the qubit implementation by using superconducting circuit.
Scheme of the tunable qubit-qubit capacitive coupling is
proposed and demonstrated~\cite{Intswitch_sc2,SQ_transmon_5,SC_transmon_tc}.
The high fidelity qubit readout using a microwave amplifier is demonstrated~\cite{SQ_transmon_ro,
SQ_transmon_ro_2,SQ_transmon_ro_3}.
Furthermore, high fidelity (99.4\%) two-qubit gate using five qubits system
is achieved. This result is the first step toward surface code scheme~\cite{SQ_transmon_4}.
These results show a good scalability towards the realization of
generating a large scale cluster state.

The flux qubit consist of a superconducting loop containing several Josephson
junctions. 
This system has a large anharmonicity and can be well approximated to
a two-level system.
Single qubit gate operations can be realized with
high speed and reasonable fidelity~\cite{SQ_flux_5}. 
Meanwhile, the best observed coherence time is an order of $10$ $\mu s$~\cite{lifetime_fq,lifetime_fq_2}.
Furthermore, the tunable coupling schemes for two qubit gate
operations are proposed and demonstrated~\cite{TC_flux_sq,TC_flux_3jj,TC_flux_sq_2,TC_flux_3jj_2,TC_flux_flux,TC_flux_sq_3,TC_flux_flux_2,TC_flux_flux_3,TC_flux_sq_5}.
Quantum non-demolition measurement of flux qubit during the
coherence time is realized by using Josephson bifurcation amplifier~\cite{JBA,JBA2,JBA3,JBA4}.

There are two typical tunable qubit-qubit coupling schemes, inductive
coupling and capacitive coupling. 
In flux qubit system, existing schemes rely on inductive coupling
with the external magnetic field.
Several schemes of the tunable qubit-qubit inductive coupling are
proposed and demonstrated~\cite{TC_flux_sq,TC_flux_3jj,TC_flux_sq_2,TC_flux_3jj_2,TC_flux_flux,TC_flux_sq_3,TC_flux_flux_2,TC_flux_flux_3,TC_flux_sq_5}.
However, it is hard to apply magnetic field to a localized region.
Due to this property, it is difficult to achieve individual
addressability of all qubits, because magnetic field may affect not
only the target qubits but also other qubits as well.
Therefore, it is important to perform two-qubit gates without
affecting other qubits by using localized fields for scalable quantum computation.

Here, we propose a way to generate and control the Ising type interaction
between four-junction flux qubits using capacitive coupling.
By using an applied voltage, we control the interaction between flux
qubits that are connected by capacitance.  Unlike the standard
schemes, our scheme does not require to change the applied magnetic
field on the flux qubit for the control of the interaction.
This may have advantage to implement two-qubit gates on the target
qubits without affecting other qubits because applying local voltages
is typically much easier than applying local magnetic flux.
We take into account of realistic noise on this type of flux qubits,
and estimate a qubit-parameter range where one can perform fault-tolerant quantum computation.
Furthermore, we show a way to generate a two dimensional cluster state
in a scalable way.
Our result paves the way to achieve the scalable quantum computation
with superconducting flux qubits.  

The rest of this paper is organized as follows:
In Section~\label{1qubit_system}, we presents the design details of our flux
qubit and effects on a flux qubit from the change of the parameters.
In Section~\ref{capacitive_coupling}, we propose our scheme for
generating Ising type interaction between capacitively coupled
superconducting flux qubits.
Moreover, we show the relationship between coupling strength and two
types of errors caused by operation accuracy, the fluctuation of
applied voltage and timing jitter.
In Section~\ref{results}, we present the analysis of our scheme
for use in multi-qubit system. 
Additionally, we discuss how to suppress the non-nearest neighbor
interactions by changing parameters and performing $\pi$ pulses.
Furthermore, we show our procedure for generating a one and
two-dimensional cluster state using qubits on square lattice in less time.

\section{Voltage controlled $\alpha$-tunable flux qubit}
\label{1qubit_system}
\begin{figure}[htbp]
 \begin{center}
  \subfigure[]{\label{1qubit_circuit}
  \includegraphics[clip,width=0.4\columnwidth]{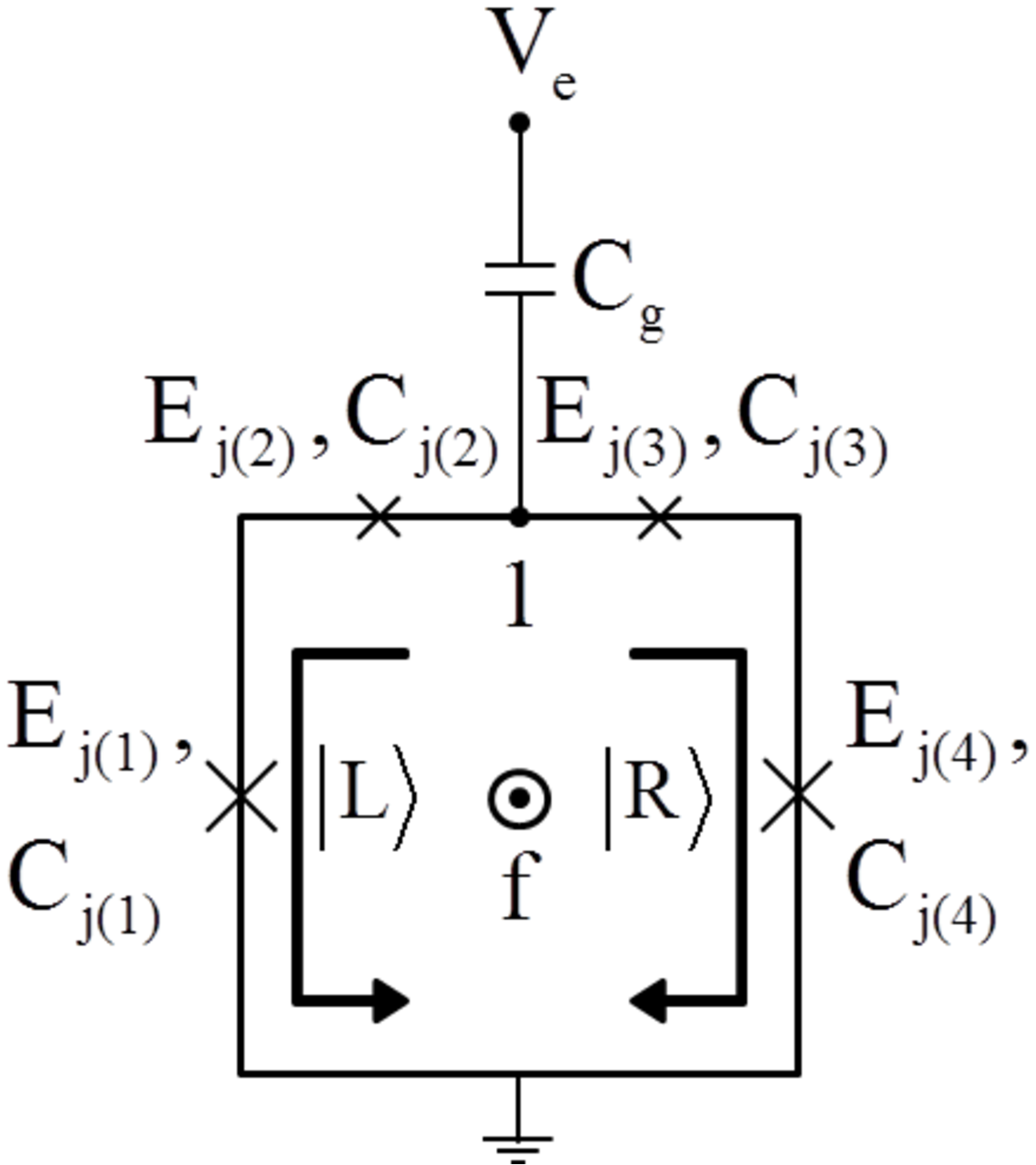}}
 \hfill
  \subfigure[]{\label{1qubit_alpha_e}
  \includegraphics[clip,width=0.56\columnwidth]{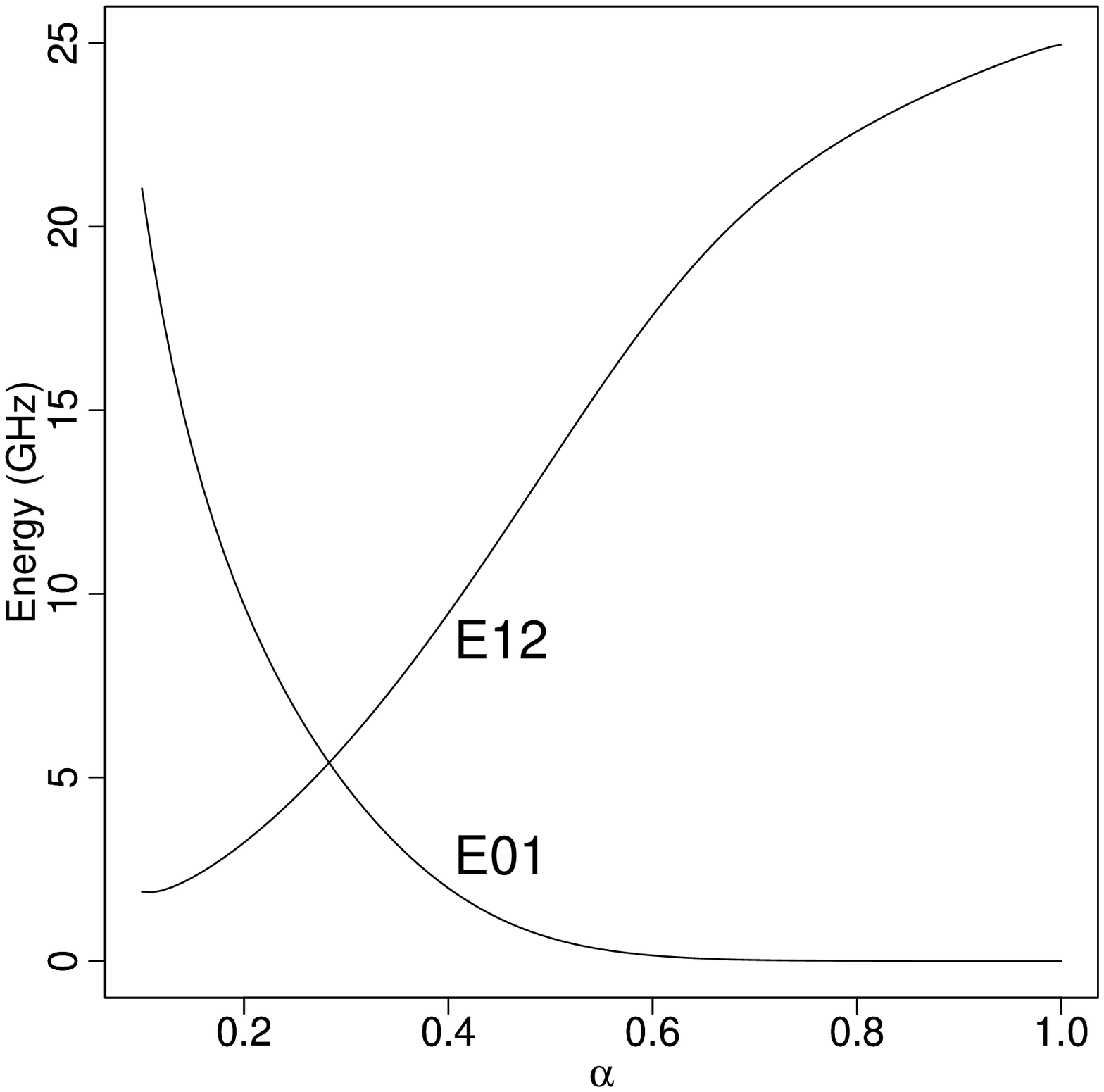}}
 \caption{(a)The circuit of a flux qubit in our
     design. This flux qubit has four Josephson junctions
     (JJ). $E_{j(n)}$ and $C_{j(n)}$ denote the Josephson energy and
     capacitance of $n$ th Josephson junction JJ$n$. The loop is
     threaded by an external magnetic flux $f$, and we can control the
     energy bias of the qubit via the magnetic flux. Node $1$
     represents the superconducting island.
   The electric potential of node $1$ is $V_{i}$. (b)
   The $\alpha$ dependence of $E_{01}$ and $E_{12}$ where $E_{01}$
   denotes energy difference between the ground state and the first
   excited state, $E_{12}$ denotes energy difference between the first
   excited state and the second excited state. Here, we set $E_{j(1)}=E_{j(4)}=200$ GHz,
   $E_{j(2)}=E_{j(3)}=40$ GHz, and $E_{j(k)}/E_{c(k)}=80(k=1,2,..,4)$.} 
 \end{center}
\end{figure}

Let us first show the circuit of a flux qubit that we propose
in Fig.~\ref{1qubit_circuit}. 
Here, X-shaped crosses denote Josephson junctions (JJ). The first
Josephson junctions (JJ$1$) and the fourth Josephson junctions
(JJ$4$) both have the same Josephson energies $E_{j}$ and capacitances
$C_{j}$. The second Josephson junctions (JJ$2$) and the third
Josephson junction (JJ$3$) both have the same Josephson energies and capacitances
that are $\alpha$ times larger than those of JJ$1$ and JJ$4$. Josephson phases
$\varphi_{n}$, which is given by the gauge-invariant phase of each
JJ$n$, are subject to the following equation:
\begin{eqnarray}
\varphi_{1} + \varphi_{2} + \varphi_{3} + \varphi_{4}=-2\pi f
\end{eqnarray}
 due to
fluxoid quantization around the loop containing phases of Josephson junctions.
$f$ denotes the external magnetic flux through the loop of the qubit in
units of the magnetic flux quantum $\Phi_{0}=\frac{h}{2e}$.
The total Josephson energy $U$ can be described as
follows:
\begin{eqnarray}
{U} &=& \sum^{4}_{k=1}E_{j(k)}\left(1-\cos \varphi_{k}\right).
\end{eqnarray}
The total electric energy $T$ can be described as follows:
\begin{eqnarray}
T &=& \frac{1}{2} \sum_{k=1}^{4}{C}_{j(k)}\left(\frac{\Phi_{0}}{2\pi}\dot{\varphi}_{k}\right)^{2}
+\frac{1}{2} C_{g}\left(V_{e}-V_{i}\right)^{2} 
\end{eqnarray}
where $C_{g}$, $V_{e}$ and $V_{i}$  denotes the
  capacitance of the gate capacitor, applied external voltage and the
  electric potential of node $1$, respectively. Here, node $1$
  represents the superconducting island.

Although the system Hamiltonian $H$ has many energy levels,
the system can be described as a two-level system (qubit) due
to a strong anharmonicity by choosing suitable $\alpha$.
We show the $\alpha$ dependence of the energy of this system
Fig.~\ref{1qubit_alpha_e}, where $E_{01}$ ($E_{12}$) denotes the
energy splitting between the ground (first excited) and the first
excited (second excited) state.
This clearly shows that system has an anharmonicity so that we can
control only the ground state and first excited state by using frequency selectivity.

$\lvert g \rangle$ and $\lvert e \rangle$ are the ground and the
first excited state of the system Hamiltonian $H=T+U$ for $f=0.5$.
In this regime, the ground state and the first excited
state of this system contains a superposition of clockwise and
anticlockwise persistent currents.
Here, $\lvert L \rangle = \frac{1}{\sqrt 2}(\lvert g \rangle +
\lvert e \rangle)$ corresponds anticlockwise persistent current and $\lvert
R \rangle = \frac{1}{\sqrt 2}(\lvert e \rangle - \lvert g \rangle)$ corresponds
clockwise one.
 
While $f$ is around 0.5, due to the anharmonicity, we can consider only
the ground state and first excited state in the Hamiltonian $H$, and
so we can simplify the $H$ into $H_{ge}$ spanned by $\lvert g
\rangle$ and $\lvert e \rangle$ as follows:  
\begin{eqnarray} 
H_{ge} &=& \frac{1}{2}\left(\Delta\sigma_{Z}+\varepsilon\sigma_{Y}\right)
\end{eqnarray}
where $\sigma_{Z}=\lvert e \rangle \langle e \rvert - \lvert g
\rangle \langle g \rvert$  and $\sigma_{Y}=-i\lvert e \rangle \langle g
\rvert + i\lvert g \rangle \langle e \rvert$ are Pauli matrices, 
$\Delta$ denotes the tunneling energy between $\lvert L \rangle$ and
$\lvert R \rangle$,
$\varepsilon$ denotes the energy bias between $\lvert L \rangle$ and
$\lvert R \rangle$.
The energy of the qubit is described as $E_{01}=\sqrt{\varepsilon^2 + \Delta^2}$.

\begin{figure}[htbp]
\includegraphics[width=86mm]{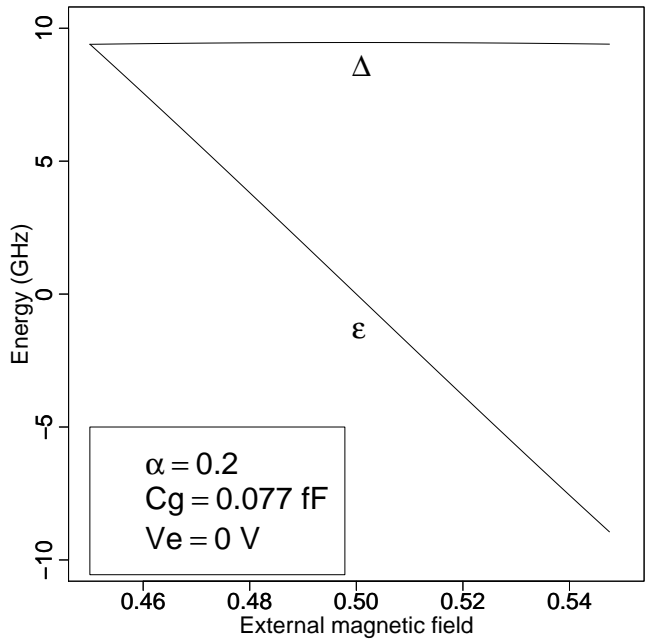}
\caption{\label{1qubit_mag}The tunneling energy
    $\Delta$ and the energy bias $\varepsilon$ against the magnetic
    flux $f$. $\varepsilon$ decreases monotonically as we increase
    $f$, while $\Delta$ is almost independent of $f$.}
\end{figure}

\begin{figure}[htbp]
 \begin{center}
  \subfigure[]{\label{1qubit_mag_e}
  \includegraphics[clip,width=0.481\columnwidth]{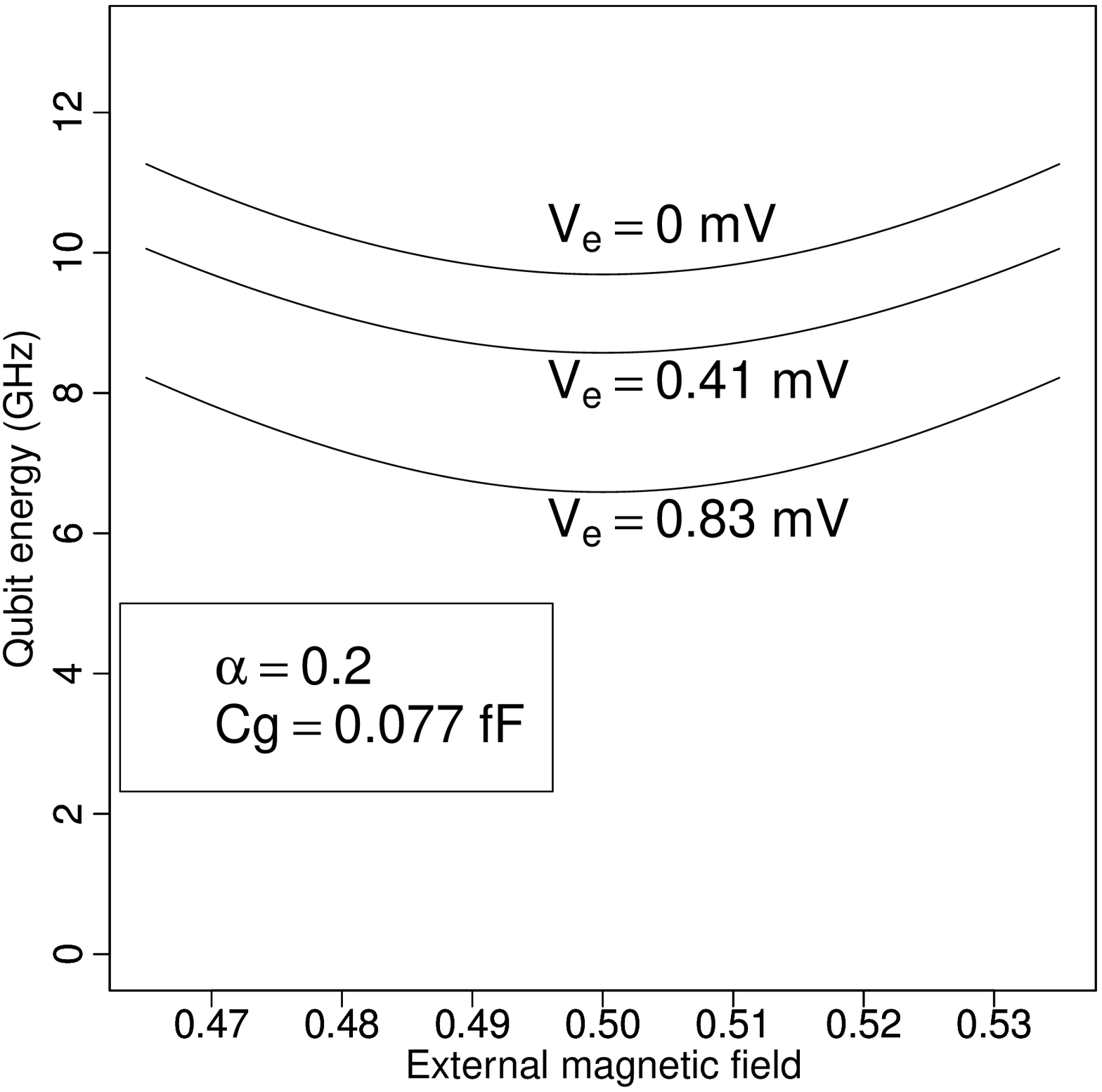}}
 \hfill
  \subfigure[]{\label{1qubit_mag_e012}
  \includegraphics[clip,width=0.481\columnwidth]{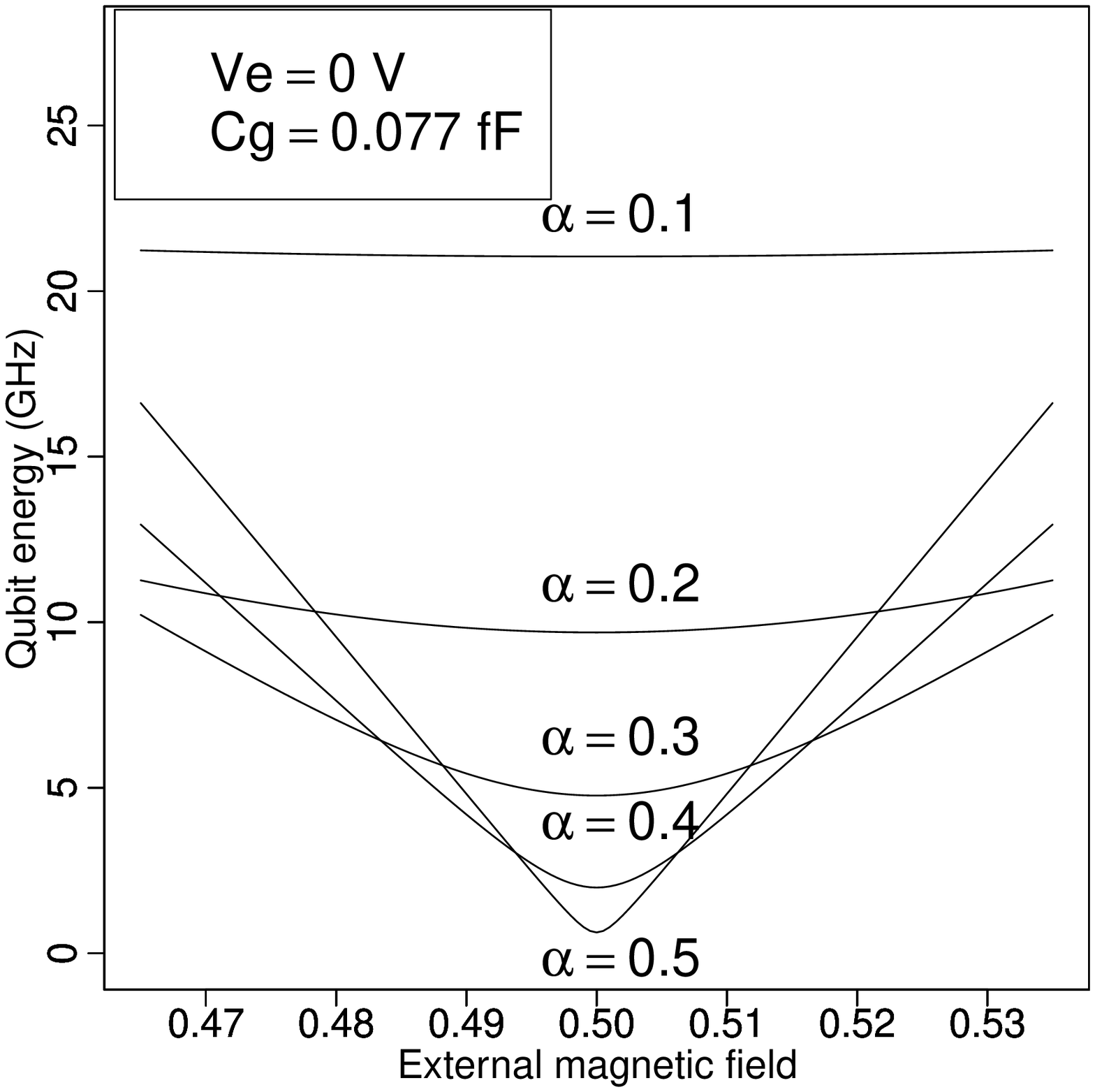}}
 \caption{~\label{1qubit_vol}(a)The relationship
     between the external magnetic flux $f$ and the energy of the qubit
     $E_{01}$ with different voltage levels. Here, we set the gate capacitance $Cg=0.16$ fF. (b) The relationship
     between $f$ and $E_{01}$ with different $\alpha$. Here, we set the gate capacitance $Cg=0.16$ fF.}
 \end{center}
\end{figure}

In this paper, unless indicated otherwise, we fix parameters as $\alpha=0.2$ and
$E_{j(1)}=200$ GHz and $E_{j(k)}/E_{c(k)}$ ratio is $80$. Here,
$E_{c(k)}=e^{2}/2C_{j(k)}$ is charge energy of each Josephson junction.
In this parameter regime, $E_{01}$ is about three times larger than $E_{12}$
as shown in Fig.~\ref{1qubit_alpha_e} so that we could consider this system as an
effective two-level system.
When $f$ is set to be near $0.5$, the derivative of the qubit energy against
the magnetic flux $\lvert\frac{d E_{01}}{df}\rvert$ takes the minimum value, so that
the qubit should be well decoupled from flux noise,
and we achieve the maximum coherent times. We call this regime
``optimal point''.
On the other hand, we can control the value of $\varepsilon$ by changing
the value of $f$.
When the energy bias $\varepsilon$ is much larger than the tunneling
energy $\Delta$, the persistent current states are the eigenvectors of
the Hamiltonian so that we can read out the qubit state with SQUID\cite{SQ_flux_2}
in $\{\lvert L \rangle , \lvert R  \rangle \}$ base.
Here we show the dependence of $\varepsilon$ and $\Delta$ against
magnetic field with no bias voltage applied in Fig.~\ref{1qubit_mag}.
It is worth mentioning that we can
control the energy of the qubit by tuning the applied voltage
$V_{e}$ while operating at the optimal point.
We show the relationship between $\Delta$ and $f$ with several values of $V_{e}$
in Fig.~\ref{1qubit_vol}.

\section{Ising type interaction using capacitive coupling}
\label{capacitive_coupling}
\subsection{Generating interaction between two-qubit system}
In this section, we show how to generate Ising type interaction
using charge coupling for superconducting flux qubit.
As a novel feature of our scheme, we use only external voltages to
switch on and off the interaction between two flux qubits.
Unlike previous schemes, external magnetic field is not required to
control the interaction in our scheme.
Since the voltage can be applied locally compared with the case of
applying magnetic field, we may
  have advantage in this scheme for scalability due to
better individual addressability when we try to control individual
  qubits.

\begin{figure}[htbp]
\includegraphics[width=86mm]{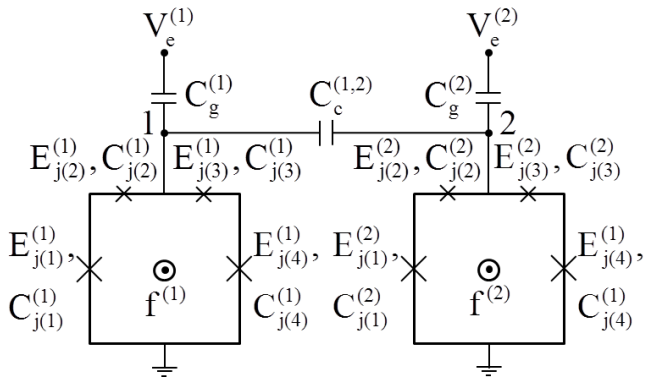}
\caption{\label{2qubit_circuit}Two flux qubits $1$, $2$ are coupled via
  capacitance $Cc^{(1,2)}$. Each flux qubit is threaded by an external
  magnetic flux $f^{(l)}$, and we can control the energy bias of the
  qubit via the magnetic flux. Node $1$ and node $2$ represent the
  superconducting islands.
  JJ$2$ and JJ$3$ at each qubit
  have the same Josephson energies and capacitances that are $\alpha$ times
  larger than those of all remaining Josephson junctions. 
  The electric potential of the island include node 1 (2) is $V_{i}^{(1)}$ $(V_{i}^{(2)})$. }
\end{figure}

Here, we show the circuit for our scheme using two superconducting flux qubits in Fig.~\ref{2qubit_circuit}.
The structure of each qubit is the same as that shown in Fig.~\ref{1qubit_circuit}.
When we apply an external voltage $V_e^{(l)}$ on each qubit, the qubit interact
with each other across the capacitor $C_{c}^{(1,2)}$.
We describe the details of this circuit in the following subsections.

\subsection{Hamiltonian}
We now consider the electric energy and potential energy of the circuit in
Fig.~\ref{2qubit_circuit} as follows: 
\begin{eqnarray}
T &=& \frac{1}{2}\sum_{k=1}^{4}\sum_{l=1}^{2}{C}^{(l)}_{j(k)}
\left(\frac{\Phi_{0}}{2\pi}\dot{\varphi}^{(l)}_{k}\right)^{2}\nonumber\\  
&&+\frac{1}{2} \sum_{l=1}^{2}{C}_{g}^{(l)}\left(V_{e}^{(l)}-V_{i}^{(l)}\right)^{2}\nonumber\\
&&+\frac{1}{2} C_{c}^{(1,2)}\left(V_{i}^{(1)}-V_{i}^{(2)}\right)^{2}\\
U &=& \sum_{k=1}^{4}\sum_{l=1}^{2}{E}^{(l)}_{j(k)}\left(1-\cos
\varphi^{(l)}_{k}\right)\\
H_{total}&=& T+U=\sum_{l=1}^{2} H_{A}^{(l)}+H_{B}\\
H_{A}^{(l)}&=&  \frac{1}{2}
\sum_{k=1}^{4}{C}^{(l)}_{j(k)}\left(\frac{\Phi_{0}}{2\pi}\dot{\varphi}^{(l)}_{k}\right)^{2}
+\frac{1}{2}{C}_{g}^{(l)}\left(V_{e}^{(l)}-V_{i}^{(l)}\right)^{2}\nonumber \\
&&+ \sum_{k=1}^{4}{E}^{(l)}_{j(k)}\left(1-\cos\varphi^{(l)}_{k}\right)\\
H_{B}&=& \frac{1}{2} C_{c}^{(1,2)}\left(V_{i}^{(1)}-V_{i}^{(2)}\right)^{2} 
\end{eqnarray}
where $C_g^{(l)}$, $f^{(l)}$, $V_{e}^{(l)}$, and $V_{i}^{(l)}$
denotes gate capacitance, external magnetic flux, applied external
voltage, and the electric potential of the island including node $l$
for the $l$ th qubit respectively. Here, node $l$ represents the
superconducting islands.

\begin{figure}[htbp]
 \begin{center}
  \subfigure[]{\label{2qubit_v_E}
  \includegraphics[clip,width=0.481\columnwidth]{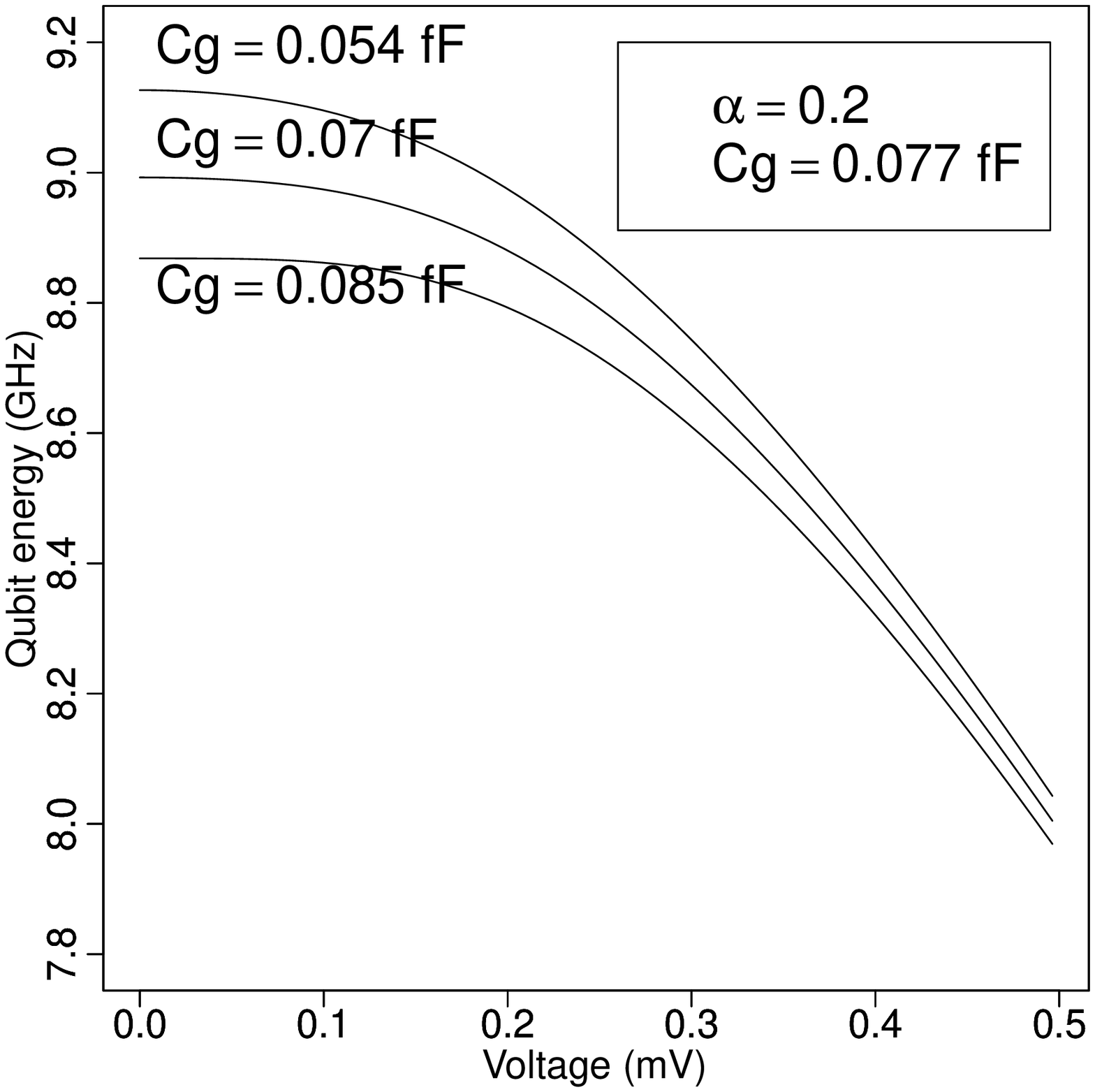}}
 \hfill
  \subfigure[]{\label{2qubit_v_g}
  \includegraphics[clip,width=0.481\columnwidth]{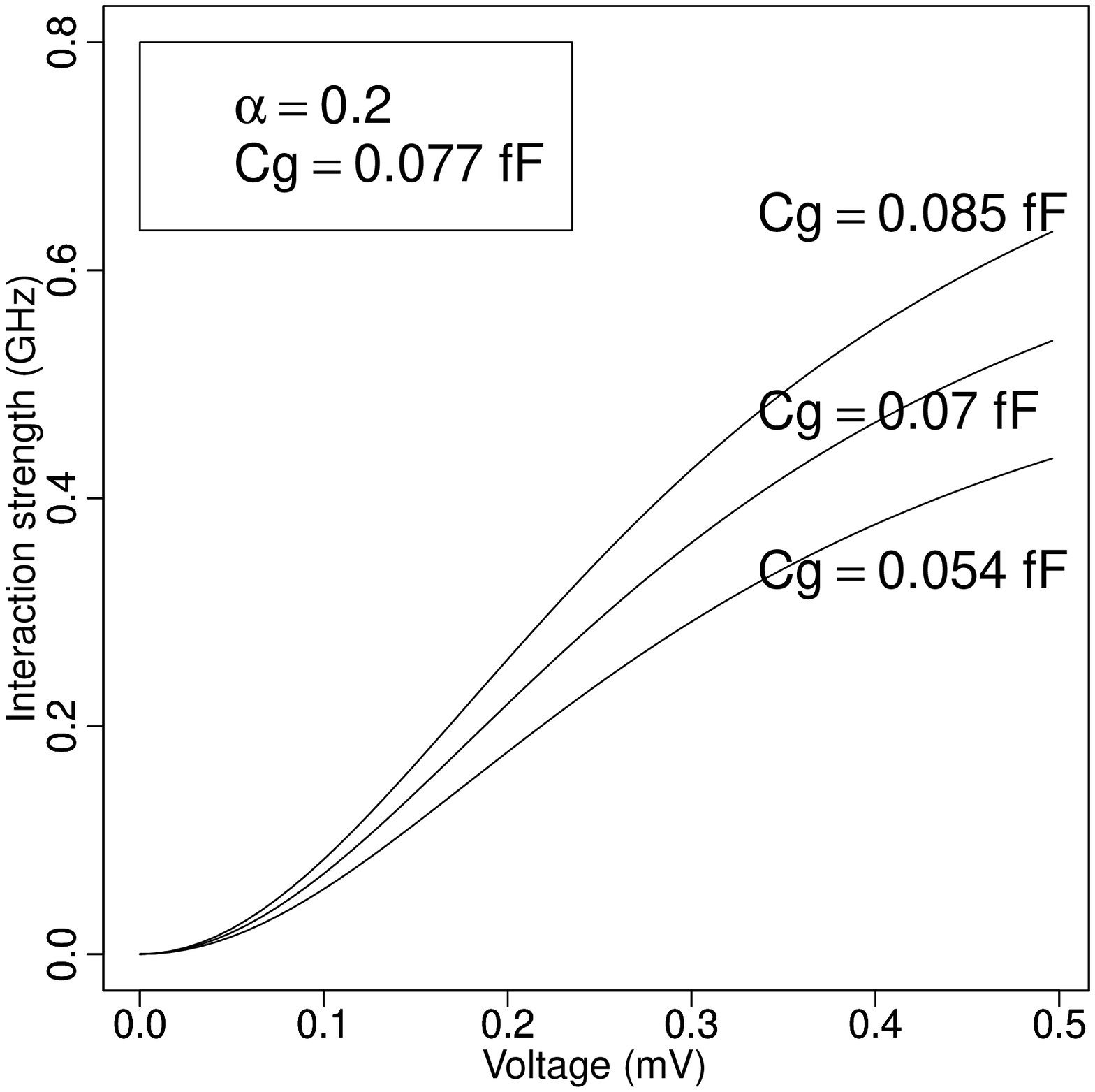}}
\caption{\label{2qubit_v_gE}The voltage dependence of the qubit
  energy $\Delta$ and the interaction strength $g$ between two qubits
  of the circuit in Fig.~\ref{2qubit_circuit}. Here, both of the gate
  capacitance $C_{g}^{l}=0.077$ fF, coupling capacitance $C_{c}=0.077$ fF.}
 \end{center}
\end{figure}

For an arbitrary $f$, we can derive the effective four-level Hamiltonian
$\hat{H}_{ge}$ of the eigenspace spanned by $\lvert g_l \rangle$ and
$\lvert e_l \rangle$ from $H_{total}$.
Here, $\lvert g_l \rangle$ and $\lvert e_l \rangle$ correspond to the
ground state and first excited state of the $l^{th}$ qubit without
interactions for $f^{(1)}=f^{(2)}==0.5$.
We expand $H_{total}$ by $\lvert g \rangle_{l}$ and $\lvert e \rangle_{l}$.
The effective Hamiltonian $\hat{H}_{ge}$ becomes as follows: 
\begin{eqnarray}
\label{eq_Ising}
\hat{H}_{ge}&=&\sum_{v_1 \in (g_1,e_1),v_2 \in
  (g_2,e_2)}\lvert v_1  v_2 \rangle \langle v_1  v_2 \rvert
\hat{H}_{total}\lvert v_1 v_2 \rangle \langle v_1 v_2\rvert\nonumber\\
&=& \sum_{l=1}^{2}\frac{1}{2}\left(\Delta^{(l)}\sigma_{Z}^{(l)}+\varepsilon^{(l)}\sigma_{Y}^{(l)}\right)+g\sigma_{Z}^{(1)}\sigma_{Z}^{(2)}
\end{eqnarray}
where $g$ denotes the Ising type interaction strength between
qubit $1$ and $2$.
We show the change of the qubit energy $E_{01}$ and the interaction
strength $g$ as a function of applied voltages in
Fig.~\ref{2qubit_v_gE}.
Large interaction strength and small derivative of qubit energy
against voltage can be achieved by the large coupling capacitance $Cc$
between each qubits.
This seems to show that one can suppress errors by increasing $Cc$.
We discuss about the errors during controlled-phase gate
operation in following section.

\subsection{Effects on interaction from change in electric field}
\label{dep_error}
To evaluate the performance of our scheme, we focus on two types of errors.
Firstly, we analyze the dephasing errors due to the fluctuations
of applied voltage. 
We define this type of error $\epsilon_{d}$ and dephasing time 
$T_{2}$ as follows: 
\begin{eqnarray}
\epsilon_{d}&=&\frac{t_{cp}}{T_{2}}, t_{cp}=\frac{\pi}{4g},
T_{2}=\frac{1}{|\frac{d E_{01}}{dv}|\delta v}
\end{eqnarray}
where we assume $t_{cp}<<T_{2}$.
Here, $t_{cp}$ denotes the necessary time to perform a
controlled-phase gate with Ising type interaction, $v$ denotes the external
voltage of each qubit, and  $\delta v$ denotes the fluctuation width of $v$.
It is  worth mentioning that $\epsilon_{d}$ has a linear relationship with $\delta v$.
To make $\epsilon_{d}$ smaller, We should obtain a parameter set where the absolute
value of the gradient of the qubit energy $E_{01}$ is small and  
the interaction strength $g$ is large.

\begin{figure}[htbp]
  \includegraphics[width=86mm]{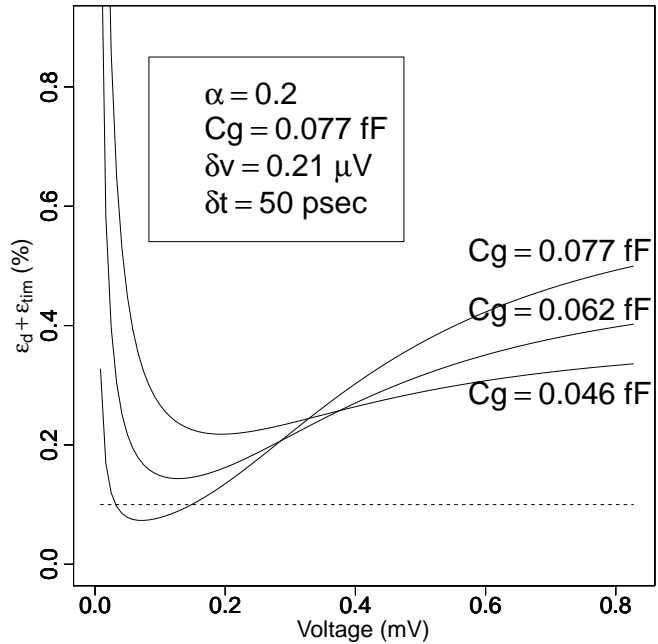}
 \caption{\label{2qubit_exvol}The total local error
   $\epsilon_{loc}(=\epsilon_{d}+\epsilon_{tim})$ as a function of voltage with
   different coupling capacitance $Cc$. Here, we set the fluctuation width of
   voltage $\delta v =0.21$ $\mu$V and the timing jitter $\delta t=50$
   psec. Dashed line denotes an error of $0.1$\%.}
\end{figure}

Secondly, we investigate the jitter error of a two-qubit gate operation.
The Ising type interaction can implement the controlled-phase gate
\begin{eqnarray}
U_{CZ}^{(1,2)}(t)&=& \exp\left(-i4gt\frac{1+\sigma_{Z}^{(1)}}{2}\frac{1+\sigma_{Z}^{(2)}}{2}\right),
\end{eqnarray}
where $g$ denotes the interaction strength in Eq.~(\ref{eq_Ising}),
$t=\frac{\pi}{4g}$ denotes the time to apply voltages,
and $U_{CZ}^{(1,2)}$ denotes a controlled-phase
gate between qubit $1$ and $2$.
By performing the controlled-phase gate on two qubits which are
initialized to $\lvert ++ \rangle_{12}$ state, we can obtain the two-qubit
cluster state.
But, the applied voltages may not create the desired state due to
error in the timing $t'=t+\delta t$, where $\delta t$ is timing
jitter.
We introduce the controlled-phase gate $U_{CZ}^{(1,2)}(t)$ including the
timing error to calculate a gate fidelity $F_{CZ}=|\langle \phi
\lvert\phi' \rangle|^{2}$ with 
\begin{eqnarray}
\lvert \phi \rangle = U_{CZ}^{(1,2)}(t)\lvert ++ \rangle,
\lvert \phi' \rangle = U_{CZ}^{(1,2)}(t')\lvert ++ \rangle.
\end{eqnarray}
Here, we define the timing error $\epsilon_{tim}=1-F_{CZ}$, and the
local error $\epsilon_{loc}(=\epsilon_{d}+\epsilon_{tim})$.
We show the $\epsilon_{loc}$ against the
applied voltage $V_{e}$ with the particular values of $Cc$ in
Fig.~\ref{2qubit_exvol}.
The threshold of local errors for fault-tolerant quantum computation
is known to be around 1\%. Also, it is known that, if the error rate is 
close to the threshold, the necessary number of qubits for the
computation drastically increases~\cite{TQC_1,TQC_2}.
Therefore, we set the threshold to $\epsilon_{loc}=0.1$\%. 
As shown in Fig.~\ref{2qubit_v_gE}, 
we can increase the coupling strength $g$ by increasing $Cc$.
Meanwhile, the strong coupling strength causes the large timing error.
Therefore, as shown in Fig.~\ref{2qubit_exvol}, the optimal voltage
exists for each of the $Cc$ which minimizes the total local error. 
In addition, by increasing $Cc$, the total error tends to be smaller.
This result shows that the large $Cc$ has an advantage for quantum
error correction against local errors.
However, for multi-qubit systems, increasing $Cc$ causes a different
problem. 
Unwanted interaction strength between non-nearest neighbor qubits increases due to the
large $Cc$. For this reason, the $Cc$ should be set to be around
$0.075$ fF. The detail of this will be discussed in Section~\ref{nnn_error}.

\section{Multi-qubit system}
\label{results}
In this section, we  generalize our scheme to multi-qubit system.
Firstly, we discuss how to control the capacitive interactions
between superconducting flux qubits via applied voltage.
Secondly, we show how to apply our scheme to generate a two
dimensional cluster state using superconducting flux qubits arranged on square lattice.

\subsection{Generating interaction between multi-qubits system}
\label{multi-qubit_system}
\begin{figure}[htbp]
\includegraphics[width=86mm]{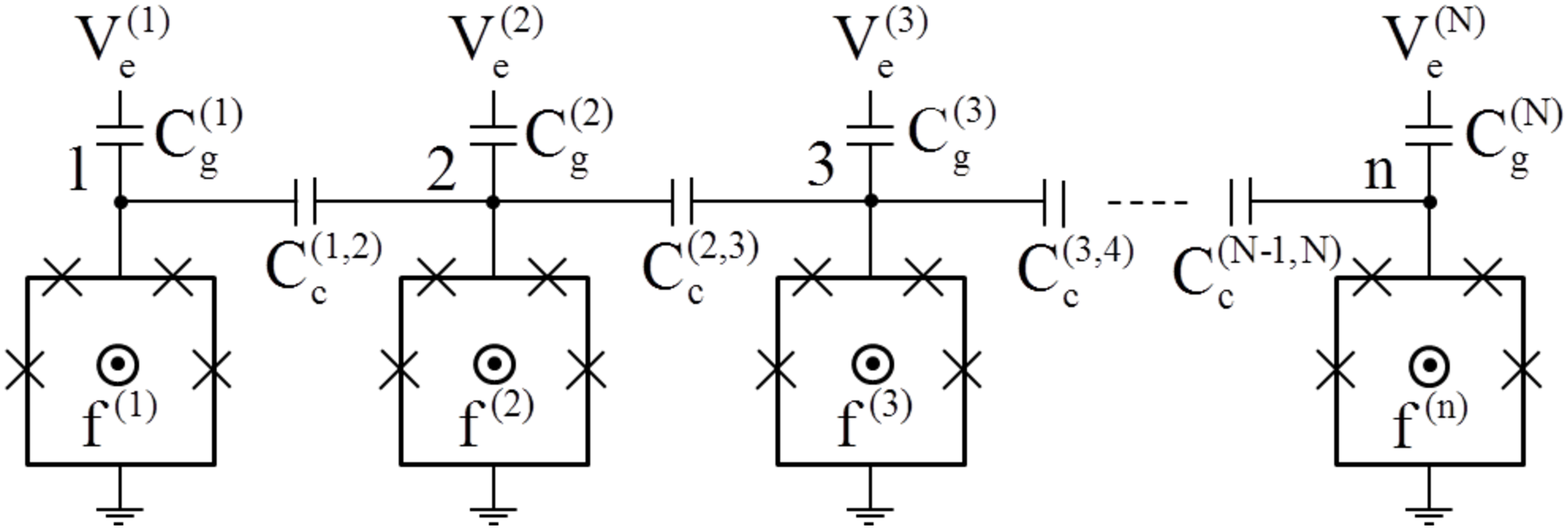}
\caption{\label{1D_circuit}A flux qubit at the
    site $j(1<j<N)$ couples with the nearest neighbor qubits via capacitance
    $C_c^{(j,j\pm 1)}$. For simplicity, we assume homogeneous flux qubits. Each
    node $j$ represents the superconducting islands. Each qubit has
    four Josephson junctions. Two Josephson
    junctions directly connected to the node have the Josephson energies and
  capacitances that are $\alpha$ times larger than the other two Josephson junctions.}
\end{figure}

Here, we discuss the interactions between capacitively coupled $N$ flux qubits
that are arranged in one dimensional line as shown in Fig.~\ref{1D_circuit}.
For simplicity, we assume homogeneous flux qubits.
$f^{(j)}$ denotes the external magnetic flux through the loop of the
$j$ th qubit.
When all flux $f^{(j)}$ are $0.5$, the system Hamiltonian is described as follows.
\begin{eqnarray}
\label{1d_hamiltonian}
\hat{H}&=&\sum_{l=1}^{N}\frac{1}{2}\Delta_{(l)}\sigma_{Z}^{(l)}
+\sum_{l,l'=1}^{N}g_{(|l-l'|)}\sigma_{Z}^{(l)}\sigma_{Z}^{(l')} 
\end{eqnarray}
where $\Delta_{(l)}$ denotes the energy of the $l$ th qubit, $g_{(|l-l'|)}$
denotes the interaction strength between each pair of qubits at a site $(l,l')$,
and $|l-l'|$ denotes  the site distance between these qubits
(e.g. when qubit $l$ and $l'$ are nearest neighbor pair, $|l-l'|=1$.).

\subsection{Generation of a one dimensional cluster state}
\label{nnn_error}
\begin{figure}[htbp]
\includegraphics[width=86mm]{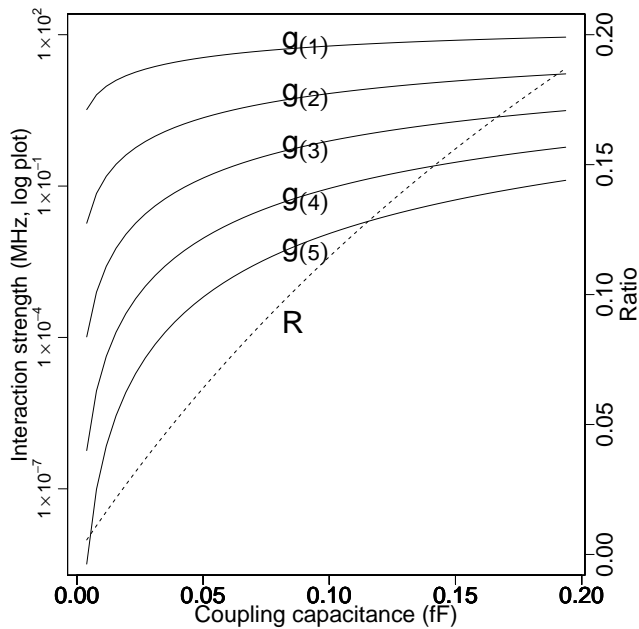}
\caption{\label{6qubit_log}The $Cc$ dependence of the interaction
  strengths and the coupling ratio
  $R\left(=\frac{g_{(2)}}{g_{(1)}}\right)$ where $g(|l-l'|)$ denotes
  the interaction strength between a pair of qubits at a site $(l-l')$.}
\end{figure}

Non-nearest neighbor interactions cause spatially-correlated errors that are
difficult to correct by quantum error correction.
 In this subsection, we show the way to evaluate
this error.
We define the ratio between nearest neighbor interaction
$g(=g_{(1)})$ and next-nearest neighbor interaction $g_{(2)}$ as
$R\left(=\frac{g_{(2)}}{g_{(1)}}\right)$ where all
  qubits are applied voltage $V_{e}$.
We show that the interaction strength $g(|l-l'|)$ decreases exponentially as
the site distance $|l-l'|$ increases, and the Ratio $R$ depends on the
coupling capacitance $Cc$ between each qubit. 
This is a striking feature in our scheme using voltage for the control
of the qubit-interaction, because the effect from any control lines
can decrease only polynomially against the site distance if one uses
magnetic field for the control.
We show the interaction strengths of 6 qubits system as a function of $Cc$
in Fig.~\ref{6qubit_log}.

If we apply voltage on all qubits, interaction occurs between such qubits.
The total error $\epsilon_{non}^{(j)}$ caused by non-nearest neighbor
interactions on $j$ th
qubit during controlled-phase operation is calculated as follows: 
\begin{eqnarray}
\epsilon_{non}^{(j)}=\sum_{n=2}^{N/2} g_{(n)}t_{cp}m_{(n)}=\sum_{n=2}^{N/2} \frac{\pi}{4} R^{(n-1)}m_{(n)}
\end{eqnarray}
where $n$ denotes the site distance between the $j$ th qubit and the coupled
non-nearest neighbor qubits, $m_{(n)}^{(j)}$ denotes the number of
such non-nearest qubits.

Such the existence of the spartially-correlated error will increase the
threshold for quantum error correction~\cite{thre_scn}.
Large capacitance tends to decrease local errors as shown in
Fig.~\ref{2qubit_exvol}, while large capacitance induces more
spartially-correlated errors as shown in Fig.~\ref{6qubit_log}.
However, when we consider the spatially-correlated error,
the error threshold value of the surface code is not well studied.
Thus, we set the upper bound of the spatially-correlated error on each qubit
$\epsilon_{non}\leq \frac{1}{10000}$ which is an order of magnitude
smaller than the threshold of local error for surface coding
scheme. If this condition is satisfied, we assume that
spatially-correlated error is enough to perform a fault-tolerant
quantum computation.
When we apply voltage on all qubits to perform controlled-phase gates
to all pairs of nearest neighbor qubit, a range of values that the coupling
capacitance $Cc$ can take while satisfying the above condition is
smaller than the proper range of $Cc$ discussed in Subsection~\ref{dep_error}.
Therefore, we do not apply voltage on all qubits but apply voltage on some of
them.
We choose pairs of nearest neighbor qubits that we will apply the
voltage, and we set a site distance $p$ between the pairs.
Then, if $R$ is small enough, $\epsilon_{non}$ of each qubit is
 the following equation:
\begin{eqnarray}
\epsilon_{non}^{(j)}=\sum_{n=p}^{N/2} \frac{\pi}{4} R^{(n-1)}\leq \frac{1}{10000}
\end{eqnarray}
where $p$ is the site distance between qubits applied by voltage.

\begin{figure}[htbp]
 \begin{center}
  \subfigure[]{\label{errors_cc}
  \includegraphics[clip,width=0.48\columnwidth]{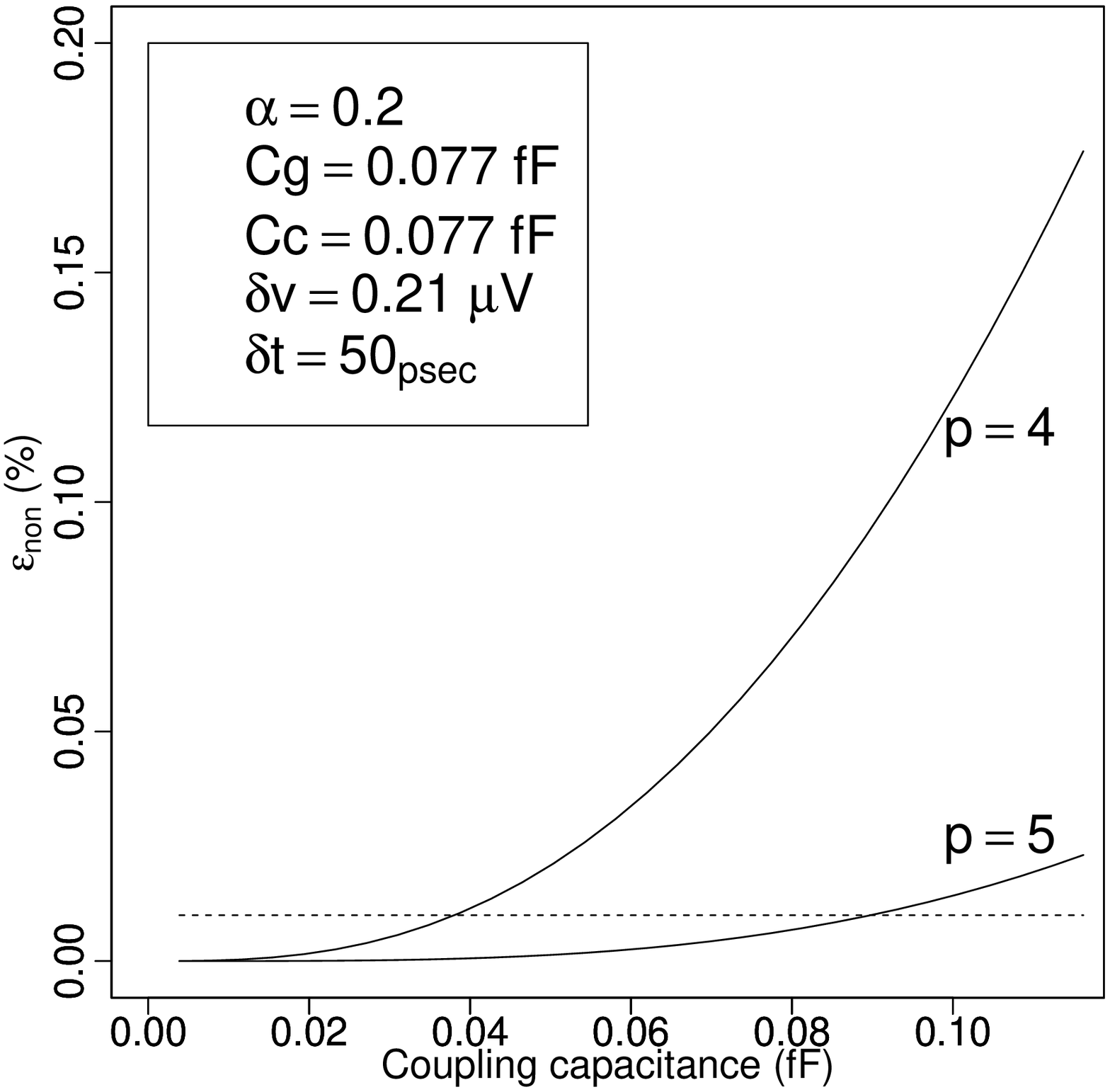}}
  \subfigure[]{\label{errors_v}
  \includegraphics[clip,width=0.48\columnwidth]{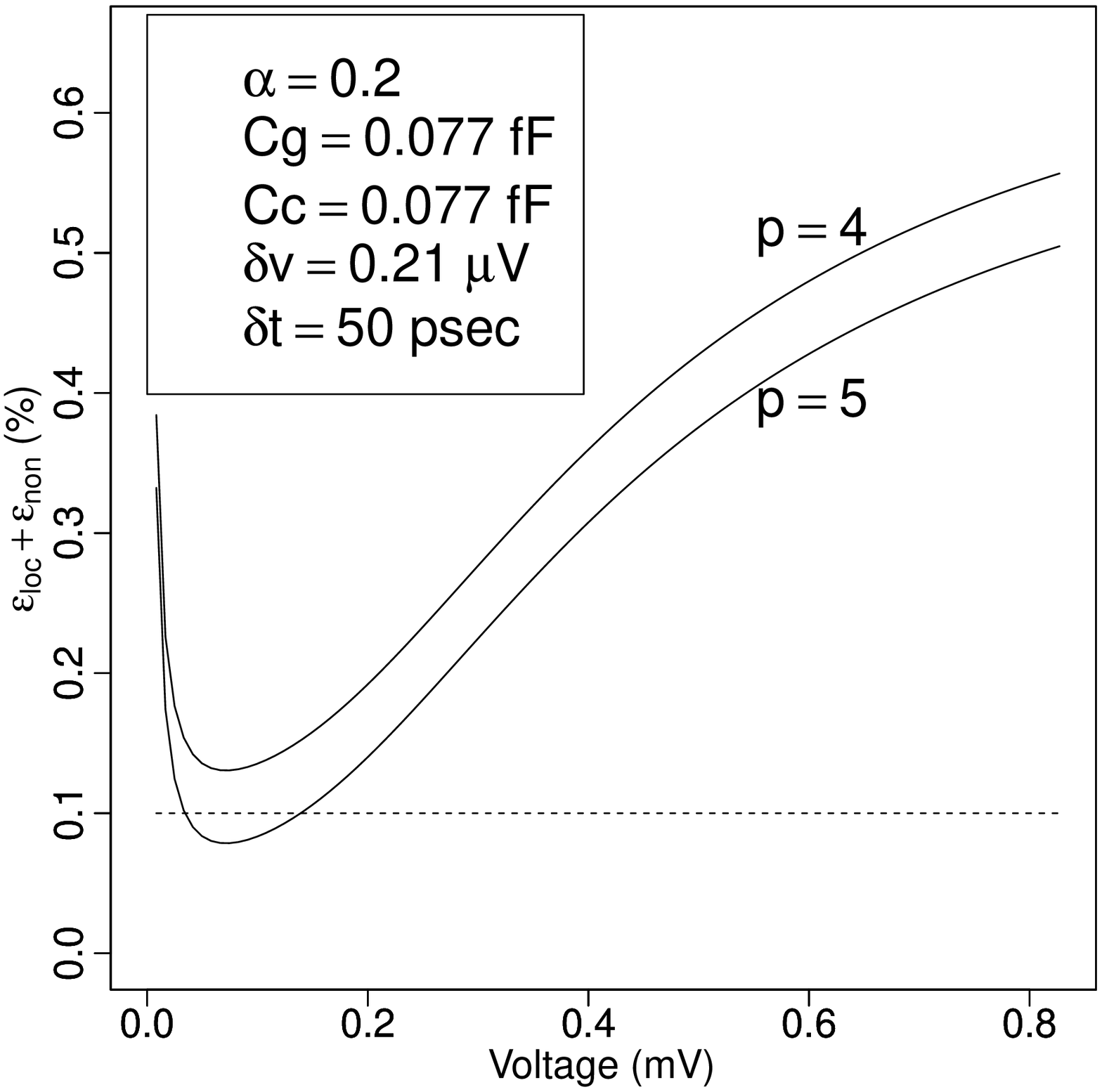}}
 \caption{\label{errors_ccv}(a)The $Cc$ dependence of the correlated
   errors. Dashed line corresponds to an error of $0.01$ \%. (b)The $V$
   dependence of the total errors. Dashed line corresponds to an error
   of $0.1$ \%.}
 \end{center}
\end{figure}

Since there are many parameters on the interaction Hamiltonian, it is
difficult to find an optimum set of parameters that minimize both of
local and spatially-correlated errors. 
Therefore, we fix the following parameters: $\alpha=0.2$,
$\delta_{v}=0.21$ $\mu$V, $\delta_{t}=50$ psec.
To determine a minimum site distance $p$ while suppressing the
correlated errors to be under $0.01$ \%, we show the $Cc$ and $V$ dependences of
the errors with $p=4,5$ in Fig.~\ref{errors_ccv}.
As shown in Fig.~\ref{errors_cc}, when $p=4$, the $\epsilon_{non}$
exceeds $0.01$ \% around $Cc=0.04$ fF. We cannot sufficiently
suppress local errors using coupling capacitance smaller than $0.07$
as shown in Fig.~\ref{2qubit_exvol}.
Thus, the site distance $p$ should be larger than $5$.
Meanwhile, when $p=5$, the $\epsilon_{non}$ exceeds $0.01$ \% around
$Cc=0.09$ fF. Then the total error of the controlled-phase operation
can be sufficiently suppressed to be less than $0.1$ \% using the
coupling capacitance $Cc$ around $0.077$ fF as shown
in Fig.~\ref{errors_v}.
Therefore, it is preferable that the site distance $p=5$ be selected.
In order to adopt sufficiently large coupling capacitance such that
the $\epsilon_{loc}$ below $0.1$ \%, we need to choose sufficiently
large $p$ such that the $\epsilon_{non}$ below $0.01$ \%.
We discuss about the way which can further reduce $p$ in the following.

\begin{figure}[htbp]
\includegraphics[width=86mm]{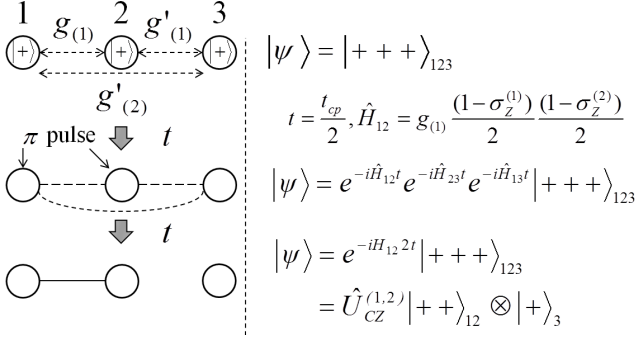}
\caption{\label{double_echo}When we perform a $\pi$ pulse on qubit
  $1$ and $2$ at $t=t_{cp}/2$, the nearest neighbor
  interaction between qubit $2$ and $3$ and the non-nearest neighbor
  interaction between qubit $1$ and $3$ are cancelled out. In such 
  way, we can perform controlled-phase gate without changing the state
of other qubits.}
\end{figure}

The $p$ determines the maximum number of controlled-phase gates that are
performed simultaneously on the same system.
For example, we can perform $\lfloor\frac{N-2}{p+1}\rfloor +1$
controlled-phase gates in parallel using $N$-qubits one dimensional system.
If we can use the  smaller $p$ without adding extra errors, we can
perform more controlled-phase gates in parallel.
So that we can generate a cluster state within a shorter operating
time.
For this purpose, we introduce the spin echo technique where
implementation of a $\pi$ pulse
(single qubit $\sigma_{X}$ rotation) 
to the target qubit could refocus the dynamics of the spin so that 
effects of interactions on the target qubit should be cancelled out.
We apply two $\pi$ pulses to  pairs of qubits to suppress spatially-correlated
errors. 
For example, we set three qubits in a raw and apply voltage
$V_{e}^{(n)}$ to the $n$ th qubits $(n=1,2,3)$ as shown in Fig.~\ref{double_echo},
where $V_{e}^{(1)}$ and  $V_{e}^{(2)}$ are equal, 
$V_{e}^{(3)}$ is an arbitrary voltage, and the strength of interaction
between qubit $1$ and $2$ is $g$.
We set each qubit to be prepared in $\lvert + \rangle$ state, let the state evolve
for a time $t_{cp}/2$, perform two $\pi$ pulses to qubit $1$ and $2$,
and let the state evolve for a time $t_{cp}/2$. The final state become as follows:
\begin{eqnarray}
\hat{U}\lvert +++ \rangle_{123}&=&\frac{1}{\sqrt{2}}\left(\lvert +0 \rangle_{12}+\lvert -1 \rangle_{12}\right)\otimes\lvert + \rangle_{3}.
\end{eqnarray}
Here, the interactions
$g_{(1)}\sigma_{Z}^{(2)}\sigma_{Z}^{(3)}$ and
$g_{(2)}\sigma_{Z}^{(1)}\sigma_{Z}^{(3)}$ are cancelled out due to the
$\pi$ pulses and we obtain a cluster state between qubit $1$ and $2$.

This method can be applied with the case of arbitrary number of
qubits. The general rules are follows: let us consider a pair of
qubits. If we perform $\pi$ pulses on both of qubits, the interaction
between them is not affected by these pulses. On the other hand, if we
perform $\pi$ pulse on one of them, the interaction between them is
cancelled out. These properties would be crucial for generating a
cluster state as we will describe.

For generating a large one dimensional cluster state using $N$ qubits
of the circuit in Fig.~\ref{1D_circuit}, we show the procedure as
follows:
\begin{figure}[htbp]
\centering
\includegraphics[width=1\columnwidth]{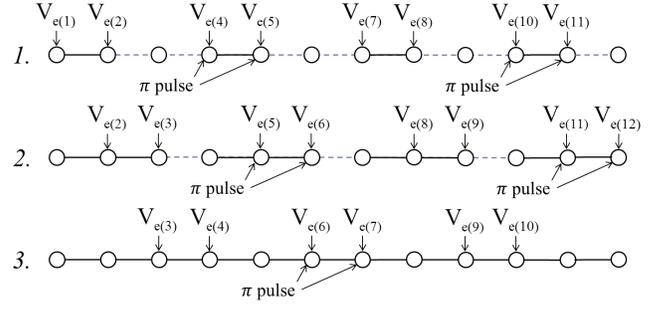}
\caption{\label{1D_cluster}The $3$-step procedure for
    generating a one dimensional cluster state. Step 1. We initialize
    $3n-2$ th and  $3n-1$ th qubits in $\lvert + \rangle$. Here,
    $n=1,2,\cdot\cdot,\lfloor\frac{N+1}{3}\rfloor$ where 
    $\lfloor x \rfloor$ is the integer part of $x$. After that we apply
    voltage on $3n-2$ th and $3n-1$ th qubits. Let the state evolve
    for a time $\frac{t_{cp}}{2}$,   perform $\pi$ pulses to $6n-2$ th
    and $6n-1$ th qubits, and let the state evolve for a time
    $\frac{t_{cp}}{2}$.  After these operations, controlled-phase
    gates have been performed between qubit $3n-2$ and $3n-1$.
    Step 2.  We initialize $3n$ th qubits in $\lvert + \rangle$. After
    that, similar to the Step 1, we perform controlled-phase gates
    between qubit $3n-1$ and $3n$. Step 3. We initialize $3n+1$ th
    qubits in $\lvert + \rangle$. After that, similar to the Step 1
    and 2, we perform controlled-phase gates between qubit $3n$ 
  and $3n+1$.}
\end{figure}
\begin{description}
\item[Step 1]
  We apply voltage to $(3n-2)$ th and $(3n-1)$ th qubit for performing
  controlled-phase gates between $(3n-2)$ th and $(3n-1)$ th qubit where
   $n=1,2,\cdot\cdot,\lfloor\frac{N+1}{3}\rfloor$.
\item[Step 2]
  We apply voltage to $(3n-1)$ th and $3n$ th qubit for performing
  controlled-phase gates between $(3n-1)$ th and $3n$ th qubit where
   $n=1,2,\cdot\cdot,\lfloor\frac{N+1}{3}\rfloor$.
\item[Step 3]
  We apply voltage to $3n$ th and $\left(3n+1\right)$ th qubit for performing
  controlled-phase gates between $(3n-1)$ th and $3n$ th qubit where
   $n=1,2,\cdot\cdot,\lfloor\frac{N+1}{3}\rfloor$.
\end{description}

\begin{figure}[htbp]
\includegraphics[width=86mm]{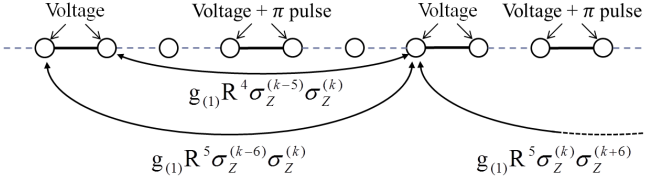}
\caption{\label{1D_echo_error}The influence of non-local interactions.
  During the controlled-phase gate, each target qubit are
  affected by non-local interactions.  We show the strength of mainly
  three non-local interactions with $k$ th qubit. These interactions
  are not cancelled out by $\pi$ pulses.}
\end{figure}

At each step of the above procedure, $\lfloor \frac{N-1}{3}\rfloor$
controlled-phase gate are performed in parallel.
At each step, we will perform the following procedure to perform the
controlled-phase gate. Firstly, prepare the qubit state in
$\lvert + \rangle$. Secondly, let the state evolve for a time
$t=\frac{t_{cp}}{2}$ according to the Hamiltonian described in Eq.~\ref{1d_hamiltonian}.
Thirdly, perform the $\pi$ pulses to suppress the non-local
interaction.  Finally, let the state evolve for a time $t=t_{cp}$.  
We show the details of these operations in Fig.~\ref{1D_cluster}
and explain how the non-local interaction is suppressed in Fig.~\ref{1D_echo_error}.
When all coupling capacitance are $Cc\leq 0.077$ fF,
the spatially-correlated error on each qubits become as follows:
\begin{eqnarray}
\epsilon_{non}^{(j)}\!=\!\sum_{n=5}^{N/2} \frac{\pi}{4}
R^{(n-1)}m_{(n)}\!\simeq\! \frac{\pi}{4} (R^{4}+2R^{5})\!\leq\! \frac {1}{10000}.
\end{eqnarray}

The $k$ th qubit is affected by mainly three non-local interactions as
shown in Fig.~\ref{1D_echo_error}. The strength of the largest
interaction is $gR^{4}$, and the strength of the other two interactions
are $gR^{5}$. The remaining non-local interactions are negligibly small.

\subsection{Generation of a two dimensional cluster state}
\begin{figure}[htbp]
  \includegraphics[width=1\columnwidth]{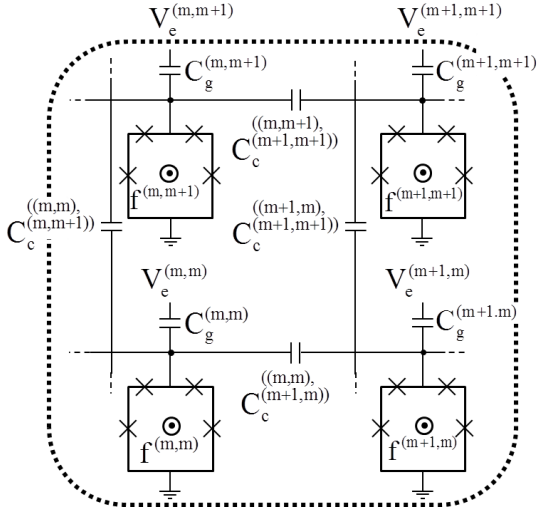}
 \caption{\label{2D_circuit}Physical circuit for generating a two
   dimensional cluster state. These four qubits correspond to the
   qubits surrounded by dot line in Fig.~\ref{2D_model}. Two Josephson junctions
   directly connected to a node (the superconducting islands) have the
   Josephson energies and capacitances that are $\alpha$ times larger than the
   other two Josephson junctions. Every flux qubit at site (j,k) couples with
   the four nearest neighbor qubits via capacitance $Cc^{((j,k)(j\pm
     1,k\pm 1))}$.}
\end{figure}

Next, we show how to generate a two dimensional cluster
state using $N^2$ flux qubits arranged on $N\times N$ square lattice. We show a
part of the circuit in Fig.~\ref{2D_circuit}.
$f^{(j,k)}$ denotes the external magnetic flux through the loop of the
qubit at site $(j,k)$. Here, $(j,k)$ corresponds to the lattice point.
When all flux $f^{(j,k)}$ are $0.5$, the system Hamiltonian is described as follows:
\begin{eqnarray}
\hat{H}_{}&=&\sum_{(l,m)}\frac{\Delta^{(l,m)}}{2}\sigma_{Z}^{(l,m)}\nonumber\\
&&+\sum_{((l,m),(l',m'))}g_{(|l-l'|+|m-m'|)}\sigma_{Z}^{(l,m)}\sigma_{Z}^{(l',m')} 
\end{eqnarray}
where $\Delta_{(l,m)}$ denotes the energy of the qubit at site $(l,m)$, $g_{(|l-l'|+|m-m'|)}$
denotes the interaction strength between each pair of qubits at site $(l,m)$
and $(l',m')$, and $|l-l'|+|m-m'|$ denotes  the site distance between
these qubits.

\begin{figure}[htbp]
  \includegraphics[width=1\columnwidth]{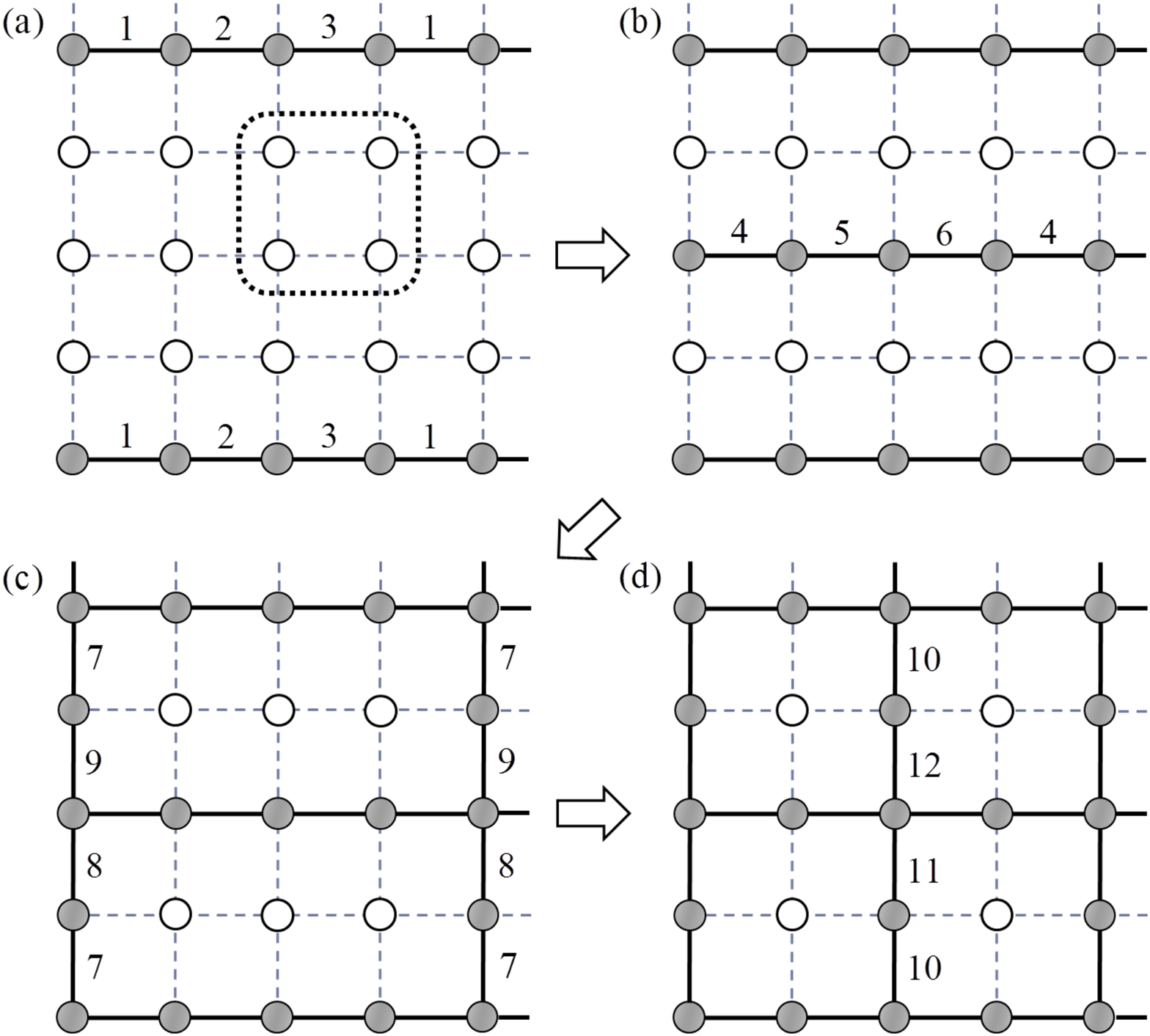}
 \caption{\label{2D_model}Schematic of our procedure for generating a
   two dimensional cluster state by graph state
     representation. Circles correspond to qubits, dashed lines
     correspond to electrically connection via a capacitance, 
     solid-lines correspond to entanglement between qubits, and
     numbers show the order in which controlled-phase gates are
     performed by our procedure. White circles
     denote separable qubit, and gray circles denote qubits constituent of cluster state(s). }
\end{figure}

Here, we show the 12-step procedure as follows for generating a two
dimensional cluster state. 
\begin{description}
\item[Step 1-3] We perform  $(N-1)\lfloor \frac{N}{4}\rfloor$
  controlled-phase gate to generate $\lfloor \frac{N}{4}\rfloor$ one dimensional
cluster states using qubits located in the $4m-3
(m=1,2,\cdot\cdot,\lfloor\frac{N+3}{4}\rfloor )$  row in the same way as
shown in Fig.~\ref{1D_cluster}. Then the spatially-correlated error of each
qubit in the $4m-3$ row is smaller than  $\frac{1}{10000}$.
We show the outline of these steps in Fig.~\ref{2D_model}(a).
\end{description}
\begin{description}
\item[Step 4-6] We perform  $(N-1)\lfloor \frac{N-2}{4}\rfloor$
  controlled-phase gate to generate $\lfloor\frac{N-2}{4}\rfloor$ one dimensional
cluster states using qubits located in the
$4p-1(p=1,2,\cdot\cdot,\lfloor\frac{N+1}{4}\rfloor )$ row in the same
way as above.
We show the outline of these steps in Fig.~\ref{2D_model}(b).
\end{description}
\begin{description}
\item[Step 7-9] We perform  $(N-1)\lfloor \frac{N}{4}\rfloor$
  controlled-phase gate to generate a two dimensional graph state as
  shown in Fig.~\ref{2D_model}(c) using qubits located in the $4m-3$ column across
$\lfloor\frac{N}{2}\rfloor$ one dimensional cluster states.
We show the outline of these steps in Fig.~\ref{2D_model}(c).
\end{description}
\begin{description}
\item[Step 10-12] We perform  $(N-1)\lfloor \frac{N-2}{4}\rfloor$
  controlled-phase gate to generate a two dimensional
cluster states using qubits located in the $4p-1$ column.
We show the outline of these steps in Fig.~\ref{2D_model}(d).
\end{description}

\begin{figure}[htbp]
\includegraphics[width=86mm]{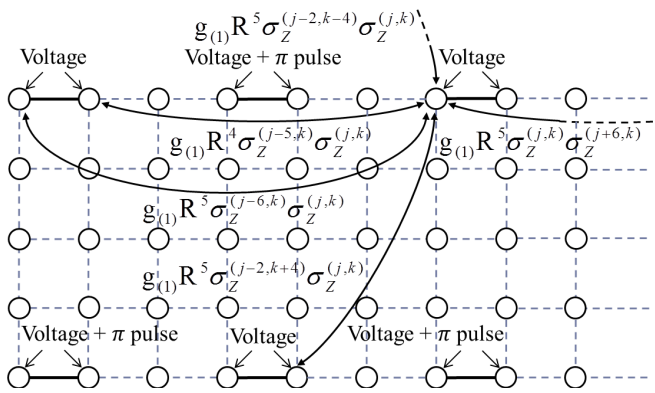}
\caption{\label{2D_cluster}Operations and the
    influence of non-local interactions in generating a two
    dimensional cluster state. In this step, we apply voltage to qubit
    at site $(3n-2,4m-3)$ and $(3n-1,4m-3)$. Let the state evolve for
    a time $\frac{t_{cp}}{2}$, perform $\pi$ pulses to qubit at site
    $(6n'-5,8m'-7)$, $(6n'-4,8m'-7)$, $(6n'-2,8m'-3)$, and
    $(6n'-1,8m'-3)$, and let the state evolve for a time 
  $\frac{t_{cp}}{2}$.  So that controlled-phase gates can be
    implemented between the pair of qubits at site $(3n-2,4m-3)$ and
    $(3n-1,4m-3)$. Here, $m=1,2,\cdot\cdot, \lfloor
  \frac{N+3}{4}\rfloor$,   $ m'=1,2,\cdot\cdot,
  \lfloor\frac{N+3}{8}\rfloor$, $n=1,2,\cdot\cdot,\lfloor
  \frac{N+1}{3}\rfloor$, and $ n'=1,2,\cdot\cdot, \lfloor
  \frac{N+1}{6}\rfloor$. Each target qubit is affected by non-local interactions from
  qubits on the same row and other rows. We show mainly five non-local
interactions with the qubit at site (j,k). These interactions are not
cancelled out by $\pi$ pulse.}
\end{figure}

We show the details of each step of above procedure for generating a 
two dimensional cluster state in Fig.~\ref{2D_cluster}.
During each step, a part of the non-local interactions are not
cancelled out by $\pi$ pulses.
When all coupling capacitance are $Cc\leq 0.077$ fF, the
spatially-correlated error on each qubits become as follows: 
\begin{eqnarray}
\epsilon_{non}^{(j,k)}\!=\!\sum_{n=5}^{N/2} \frac{\pi}{4}
R^{(n-1)}m_{(n)}\!\simeq\! \frac{\pi}{4} (R^{4}\!+\!4R^{5})\!\leq\!\frac{1}{10000}.
\end{eqnarray}
The qubit at site $(j,k)$ is affected by mainly five non-local interactions as
shown in Fig.~\ref{2D_cluster}. The strength of the largest
interaction is $g_{(1)}R^{4}$, and the strength of the other four interactions
are $g_{(1)}R^{5}$. The remaining non-local interactions are negligibly small.

\section{Conclusion}
In conclusion, we suggest a new way to generate Ising interaction
between capacitively-coupled superconducting flux qubits by using an
applied voltage, and we also show architecture about how to make a
two-dimensional cluster state in this coupling scheme.  Unlike the
standard schemes, our scheme does not require to change the applied
magnetic field on the flux qubit for the control of the
interaction. Since applying local
voltages is typically much easier than applying local magnetic flux,
the scheme described in this paper may have advantage to
perform two-qubit gates on target qubits without affecting any other qubits.
Our result paves the way for
scalable quantum computation with superconducting flux qubits.

\bibliography{201407_PRA}

\end{document}